\documentclass[12pt]{article}
\usepackage{amsmath}
\usepackage{graphicx}
\usepackage{epstopdf}
\usepackage{booktabs}
\usepackage{lscape}
\usepackage{setspace}
\usepackage{subcaption}
\usepackage{bm}
\usepackage{bbm}
\usepackage{subfloat}
\usepackage[bottom]{footmisc}
\usepackage{natbib}
\bibpunct{(}{)}{;}{a}{,}{,}
\usepackage[margin=0.83in,footskip=0.25in]{geometry}
\usepackage[capposition=top]{floatrow}
\usepackage{newtxmath}
\usepackage{longtable}
\usepackage{rotating}
\usepackage{url}

\title{The Use of Quantile Methods in Economic History\thanks{We are grateful to Roger Koenker for very useful comments, and we thank Matías Méndez Montenegro for research assistance.  We acknowledge the funding of the Chilean National Research Agency (ANID) via Proyecto Anillos ``Bienestar y Desigualdad en Chile, 1810-2020, grant SOC PIA 180001.}}
\author{Damian Clarke\thanks{Department of Economics, University of Chile \& IZA. 
Contact: dclarke@fen.uchile.cl.} \and Manuel Llorca Ja\~na\thanks{Escuela de Administración Pública, Universidad de Valpara\'iso.  Contact: manuel.llorca@uv.cl.}
\and Daniel Pailañir\thanks{Department of Economics, University of Chile.  Contact: dpailanir@fen.uchile.cl.}}
\date{\today}

\begin{document}
\begin{spacing}{1.5}
  \maketitle
  \begin{abstract}
    Quantile regression and quantile treatment effect methods are powerful econometric tools for considering economic impacts of events or variables of interest 
    beyond the mean.  The use of quantile methods allows for an examination of impacts of some independent variable over the 
    entire distribution of continuous dependent variables.  Measurement in many quantative settings in economic history have as a key input continuous outcome variables of interest.  Among many other cases, human height and demographics, economic growth, earnings and wages, and crop production are generally recorded as continuous measures, and are collected and studied by economic historians.  In this paper we describe and discuss the broad utility of quantile regression for use in research in economic history, review recent quantitive literature in the field, and provide an illustrative example of the use of these methods based on 20,000 records of human height measured across 50-plus years in the 19\textsuperscript{th} and 20\textsuperscript{th} centuries. We suggest that there is considerably more room in the literature on economic history to convincingly and productively apply quantile regression methods. 
  \end{abstract}


\clearpage

\section{Introduction}
Correctly capturing key patterns and relationships in economic history often requires the measurement of entire distributions of variables.  For example, consider human demographics.  Stunting, or very low heights owing to impaired growth in childhood, is associated with lower cognitive ability in later life, poorer health during adulthood, and reduced labour market earnings \citep{JayachandranPande2017,fogel2004,floudetal2011}.  And while stature is known to be broadly associated with health and other measures of well-being in many populations and time-periods \citep{Costa2004,VOGL201484,Deaton13232}, these relationships are not constant over the entire range of heights, nor even montonic in all populations.  Thus, if one wishes to evaluate some historical event as an input to height or other demographic measures, it is necessary to determine how the event has impacted heights across the entire distribution.  Similar phenomena are encountered in many applications in economic history, such as economic growth and its impacts on inequality in income distributions, determinants of historical salary distributions, and the study of mortality declines in higher versus lower mortality settings.  In each case, the distributional effects of policies or events can be as important as their mean effects in the population under study.

In this article we seek to motivate the importance of considering the distributional impacts of historical events, and specifically survey a broad series of methods which are ideally designed for such analyses.  In particular, we lay out a range of methods related to quantile analyses, such as quantile regression, and the estimation of quantile treatment effects.  These methods have emerged in a long line of theoretical and computational advances in the econometric literature, first fully described in \citet{KoenkerBassett1978}, and comprehensively discussed in a range of papers or books since \citep{Koenker2017,KoenkerHallock2001,Lamarche2019,koenker_2005}.  While four decades have passed since the publication of \citet{KoenkerBassett1978}'s seminal paper on quantile regression, this is still an area of active research, in particular with recent advances considering quantile treatment effects, and identification in broader circumstances. 

Despite being well-suited to applications in economic history where dependent variables of interest are often continuously distributed, these are arguably under-utilised in empirical applications.  As well as providing an overview of these methods including recent extensions which are likely to be of particular interest to practitioners in economic history, we provide a survey of their usage in economic history, an analysis of their use, and \emph{potential} for their use in papers published over the past 30 years in the field of economic history. In a survey across all principal economic history journals we find around 50 papers that have used quantile regression in some way.  However, in a deeper analysis of all the papers published in the journal \emph{Explorations in Economic History}, we find that while around 60\% of papers are based on quantitative analyses of continuous dependent variables and thus potentially suited to analysis using quantile regression or related methods, only around 1\% of papers actually apply these tools to consider effects beyond the mean.

This article thus seeks to motivate the ``use'' of quantile analyses in economic history in two ways.  The first is to describe how they can be productively used by practitioners in empirical studies to capture relationships of interest which may not be gleaned from simple mean or other average estimators.  And the second is to document how they have been used (and arguably under-used) in the literature on economic history up to this point. 

In what remains of this paper, we first provide a summary of key methods for distributional analyses in empirical methods in section \ref{scn:theory}.  In section \ref{scn:hist} we provide both an overview of papers based on quantile analysis methods in economic history, as well as a review of all papers in a highly cited economic history journal, and their suitability for methods of this type.  Finally, in section \ref{scn:example} we document a specific example based on microdata on human height covering around 20,000 individuals exposed to divergent rates of economic growth throughout their life. In closing, we make a number of points related to the computational implementation of these methods.

\section{Quantile Regression}
\label{scn:theory}
Generally, in empirical applications one wishes to consider the impact of some group of independent variables $x_k$ for $k\in\{1,\ldots,K\}$ on a specific dependent variable, denoted $y$.  Where observations $i$ refer to units (such as individuals), this is often parameterised using a linear regression model:
\begin{equation}
  \label{eqn:linMod}
  y_i = \beta_1 + \beta_2x_1 + \ldots + \beta_Kx_K + u_i,
\end{equation}
where $u_i$ is an unobserved error term.  Frequently, and indeed, in a the majority of papers in economic history (see Section \ref{scn:hist}) estimation is implemented using the ordinary least squares (OLS) method.  Mathematically, this procedure simply consists of finding the $K$ parameter estimates of $\beta$ which minimize the following problem:
\[
\widehat\beta=\arg\min_{\beta\in\mathbb{R}^K} \sum_{i=1}^N(y_i-\bm{x_i}^\prime\beta)^2,
\]
where $\bm{x_i}$ refers to the vector of values of each the independent variables $x_k$ for unit $i$.  This optimization returns a vector of parameters capturing the mean impacts of each $x_k$ on the outcome of interest $y$.  While mean impacts across the distribution of $y_i$ may be a logical summary parameter in many cases, it is not the only point with meaningful empirical content. Often, other points of the distribution of $y$ may be as, or more important, than the mean, particularly in historical processes where extreme outcomes may be of particular interest.  This suggests the importance of modelling options allowing for additional heterogeneity.\footnote{Heterogeneity can be observed for a number of reasons in empirical studies.  In this paper we are interested in heterogeneity in the impacts of some independent variable(s) over the full distribution of a dependent variable of interest.  A useful alternative example of the interest of heterogeneity in economic history is provided by \citet{BisinMoro2020} who discuss heterogeneity in the context of differential take-up of some treatment, and resulting sample-based heterogeneity due to the estimation of Local Average Treatment Efects (LATEs).}  Below we describe a range of quantile estimation techniques which allow for focus across any points of the distribution of outcomes of interest.

\subsection{The Linear Quantile Regression Model}
Quantile regression considers the full distribution of some dependent variable $y$.  Thus, at a minimum, $y$ requires a distribution with considerable variation, and is inappropriate in cases where $y$ is a binary or categorical measure. We will denote as $\tau$ the quantiles of the distribution, such that (for example) $\tau=0.5$ indicates the median of the distribution, and $\tau=0.1$ indicates the 10\textsuperscript{th} percentile, or the point of the distribution below which 10\% of the observations of $y$ are observed.  As originally documented in \citet{KoenkerBassett1978}, we can estimate the parameter vector $\beta(\tau)$ capturing the impact of each $x_k$ on the $\tau$\textsuperscript{th} quantile of $y$ based on the following minimization problem:
\begin{equation}
\label{eqn:qreg}
\widehat\beta(\tau)=\arg\min_{\beta\in\mathbb{R}^k}\left[\sum_{i:y_i\geq \bm{x_i}^\prime\beta} \tau\left|y_i-\bm{x_i}^\prime\beta\right|+\sum_{i:y_i<\bm{x_i}^\prime\beta} (1-\tau)\left|y_i-\bm{x_i}^\prime\beta\right| \right].
\end{equation}
Note that here, estimation is based on absolute deviations of $\bm{x_i}^\prime\beta$ from $y_i$ rather than quadratic distances in OLS, and indeed, in the case that $\tau=0.5$ (the median) this formula collapses to the Least Absolute Deviations estimator.  In all cases except for the median, this minimization problem uses the quantity $\tau$ to `tilt' the estimates towards data which is lower or higher in the distribution of $y$, as can be observed in the two terms within the parentheses in equation \ref{eqn:qreg}: when $\tau$ is between 0 and 0.5, more weight is given to the right-hand summation for units $i$ whose $y_i$ is less than the conditional mean $\bm{x_i}^\prime\beta$, whereas when $\tau$ is between 0.5 and 1, more weight is assigned to the left-hand summation considering units whose $y_i$ is greater than the conditional mean.


Equation \ref{eqn:qreg} is the well-known linear quantile regression model which is implemented as standard in many computational languages.  Frequently, rather than focusing on a particular quantile of interest, estimates $\widehat\beta(\tau)$ are documented over a range of quantiles in a graphical manner (refer for example to the illustration in section \ref{scn:example} of this paper.  What's more, with modern computational tools, standard errors can be calculated quite simply along the range of the distribution of $y$, which permits formal hypothesis tests and other inferential procedures at each quantile considered. Formally, following notation from \citet{Lamarche2019}, the approximate distribution of the vector $\widehat\beta(\tau)$ can be written as: \[
    \widehat\beta(\tau)\sim \mathcal{N}\left(\beta,\frac{1}{N}\tau(1-\tau)H^{-1}JH^{-1}\right),
\]
where $\mathcal{N}(\cdot)$ refers to a Gaussian distribution and the matrices   $J=\lim_{N\rightarrow\infty}\frac{1}{N}\sum_{i=1}^Nx_ix_i^\prime$, and $H = \lim_{N\rightarrow\infty}\frac{1}{N}\sum_{i=1}^Nf_i(x_i^\prime\beta(\tau))x_ix_i^\prime$, with $f_i(\cdot)$ being the conditional density function of $y$.  Note here two key implications: firstly, that this inference holds asymptotically, ie as the sample size grows, and secondly that a key `ingredient' is the estimate of the density $f_i(x_i^\prime\beta(\tau))$ at quintile $\tau$. If one is willing to assume that error terms are identical and independently distributed (iid), the $f_i$ term is identical among all units $i$, and estimation is simplified, for example using the fitted value of the density at the quantile of interest.\footnote{There are also other reasonable approches generally included as standard in statistical software, such as using a kernel density estimator or residual quantile function to estimate the density at quantile $\tau$.  These are all available in programs such as Stata with the \texttt{qreg} program, or R with the \texttt{quantreg} library.}  However, if heteroscedasticity robust estimates are desired, more complex `sandwich' estimates are necessary.  While these are also implemented as standard in computational routines in languages such as Stata or R, further background on these procedures can be found in \citet[chapter 3]{koenker_2005}.  Alternative options which avoid the estimation of these matrices consist of using bootstrap resampling methods for inference.  A review of such resampling methods in quantile regression is provided by \citet{He2018}.  Finally, note that solutions have also been proposed to resolve cases where clustered inference is desired, allowing for correlated shocks within groups, as well as heteroscedasticity \citep{Hagemann2017}, once again based on bootstrap resampling procedures.

\subsection{Quantile Treatment Effects}
\label{sscn:QTE}
While this quantile regression can be used for $x_k$ of arbitrary forms and dimensions, a particular case of interest is that of treatment effect models, with some binary `treatment' of interest, which we denote $D_i$.  Note that while quantile regression requires continuous distributions for the outcome variable of interest, there is no limit on the nature of independent variables $x_k$, including binary and categorical measures.  The standard average treatment effect aims to identify the mean impact of receiving treatment (versus not receiving treatment) on dependent variable $y$: 
$\Delta=E(y_i|D=1)-E(y_i|D=0)$.  This assumes independence of $D$ from unobservables.  Extending this average treatment effect to its quantile treatment effect (QTE) analogue gives.
\begin{equation}
    \label{eqn:QTW}
    \Delta(\tau)=Q_y(\tau|D=1)-Q_y(\tau|D=0),
\end{equation}
where $Q_y(\tau)$ refers to the value of $y$ at a particular quantile $\tau$ of the distribution of $y$.\footnote{Formally, this is $Q_y(\tau)=\inf\{y:F(y)\geq\tau\}$, where $F(\cdot)$ refers to the cumulative density function of $y$.} Such an estimate, beyond the mean is likely to be relevant for a range of historical policies, for example examining whether exposure to a specific type of economic or political system, exposure to a historical environmental shock, or exposure to a historical policy has divergent impacts across the distribution of individual outcomes capturing well-being.  All such examples refer to binary `treatment' statuses, and hence are potentially appropriate for QTE methods and their extensions discussed in the following sub-section, provided that outcomes of interest $y$ are continuous measures.  In practice, these estimates can be generated using regression following the procedure laid out in the previous section.  We return to discuss computational implementations in the final section of this paper. 


This setting can be extended considerably, for example to include covariates, and/or to explicitly consider selection into treatment.  This can be done in a number of ways, such as by attempting to correct for selection non-parametrically \citep{Bitleretal2006,Firpo2007}, or extending to panel data settings and using recent methods such as difference-in-differences \citep{CallawayLi2019}.  
A particular case where QTE methods are extended to deal with endogeneity is laid out in the following sub-section when discussing instrumental variables and Local QTEs. 
While the inclusion of covariates can complicate the QTE setup \citep{Koenker2017}, suggestions on how to deal with this have been proposed in \citet{Firpo2007,FrolichMelly2010,Callaway2018} based on propensity score methods and re-weighting techniques.  For applied work, a particularly useful reference discussing a range of practical estimation methods for quantile treatment effects is the paper of \citet{FrolichMelly2010}.

\subsection{Quantile Regression Extensions}
While standard quantile regression and QTE implementations can provide illustrative results which allow for an understanding of effects beyond the mean, recent theoretical and computation advances in these methods offer considerable additional benefits, including the possibility to address a number of issues in standard quantile regression models, and more explicitly address questions of causality.  We discuss the broad scope of these recent advances here, providing key references.  Full modeling considerations are available in the papers discussed below, however particularly useful overviews of these methods at more length can also be found in the survey articles or handbook chapters of \citet{Koenker2017} (discussing all below points), or \citet{Wei2017,Lamarche2019,MellyWuthrich2018,Chernozhukovetal2018} with focuses on particular points.

\paragraph{Models to Correct for Measurement Error}
In the case of variables collected over a considerable time-frame from diverse data sources as is often the case in studies in economic history, measurement error or issues of sample selection are valid concerns.  Fortunately, a number of methods suggest ways that measurement error can be accounted for within the framework of quantile regression (under assumptions about the nature of these errors).  For example, \citet{WeiCarroll2009} document potential estimation methods when in the presence of mesurement error in independent variables, and alternative methods are proposed by \citet{Wangetal2012}.  The crux of these models is that if rather than observing a true variable $x$ we observe some noisy proxy, we must consider the conditional \emph{expectation} of $x$ based on the noisy proxy, rather than simply using the proxy in the regression. Additionally, even if there is limited error in the measurement of variables, samples may be selected in certain ways, for example if only certain types of records survive following collection of historical data, or if data are based on suriving individuals.  \citet{ArrellanoBonhomme2017} propose methods for quantile regression under selection based on an exclusion restriction for estimation, while \citet{Blancoetal2013} (see also \citet{Lamarche2019}) propose partial identification methods as way ahead in such circumstances.  Computational implementations of such selection models are available, for example via the programs of \citet{BE2020,Siravegna2020}.  Interested readers are pointed to the handbook chapter of \citet{Wei2017} which provides a deeper discussion of these issues.

\paragraph{Endogeneity, Instrumental Variables and Causality} In many settings of interest in economic history, independent variables of interest will be endogenous---corrleated with unobserved factors which are themselves correlated with outcomes of interest.  As is the case in standard econometric models, quantile regressions or QTE models do not allow for causal inference when in the presence of endogeneity.  However, there is a rich stream of work which seeks to extend standard linear instrumental variable (IV) models into a quantile framework.  Here in particular we discuss the Local Quantile Treatment Effect (LQTE) approach, which, under a number of key assumptions provides consistent (causal) estimates for certain sub-groups of the population across the entire distribution of the outcome variable.

The LQTE framework extends the well-known Local Average Treatment Effect (LATE) framework laid out in \citet{ImbensAngrist1994}.  Here, causal estimation using an instrumental variable is based on a monotonocity assumption that requires that the instrumental variable induces exogenous variation in the dependent variable of interest which shifts all individuals weakly in the same direction, for example acting as an as-good-as-randomly assigned incentive or disincentive to uptake an endogenous variable of interest.  Here we summarise this and the other identifying assumptions, as well as how an LQTE is estimated, with a binary instrumental variable denoted $Z$.  A chapter length discussion of these methods, as well as extensions to other circumstances is provided by \citet{MellyWuthrich2018}.  Following the notation of section \ref{sscn:QTE}, consider an outcome variable of interest $y$, a binary treatment variable of interest $D$, as well as the instrumental variable $Z$.  Under 5 assumptions; instrumental independence (or validity), the exclusion restriction, instrumental relevance, monotonicity and the stable unit treatment value assumption (SUTVA) limiting interaction between treatment statuses, \citet{ImbensAngrist1994} prove that the following Wald estimator gives the LATE:
\[
\frac{E(y_i|Z_i=1)-E(y_i|Z_i=0)}{E(y_i|D_i=1)-E(y_i|D_i=0)}
\]
Note that this LATE is a consistent estimate of the impact of treatment ($D_i$) on outcomes ($y_i$) for those individuals whose treatment status would be changed by the instrument (the `compliers').

In order to extend this framework to a quantile framework, distributional results are provided by \citet{ImbensRubin1997,Abadie2002}.  These give cumulative density functions specifically for the compliers, in this case which we denote $F^c_{Y_1}$ and $F^c_{Y_0}$.  Here superscript $c$ refers to the population of compliers, and $F$ refers to the cumulative density of outcomes for cases where treatment is received ($F^c_{Y_1}$) or not received ($F^c_{Y_0}$).\footnote{In extensive form, these can be represented \citep{Abadie2002} as: 
\[
F^c_{Y_1}(y)=\frac{E[\mathbbm{1}(Y\leq y)D|Z=1]-E[\mathbbm{1}(Y\leq y)D|Z=0]}{E(D|Z=1)-E(D|Z=0)},
\]
and
\[
F^c_{Y_0}(y)=\frac{E[\mathbbm{1}(Y\leq y)(1-D)|Z=1]-E[\mathbbm{1}(Y\leq y)(1-D)|Z=0]}{E(1-D|Z=1)-E(1-D|Z=0)}. 
\]
}
Given these definitions based on an IV and compliers, it is very easy to return to the notation from section \ref{sscn:QTE}, which gives the LQTE estimator.  As a clear parallel to equation \ref{eqn:QTW}, this is defined as
\begin{equation}
    \label{eqn:LQTE}
    \Delta^c(\tau)=Q^c_{Y_1}(\tau)-Q^c_{Y_0}(\tau),
\end{equation}
where as previously, $\tau$ refers to quantiles of the relevant conditional density function above.\footnote{In the interests of completeness, this is:
\begin{eqnarray}
Q^c_{Y_1}(\tau)&=&\inf\{y:F^c_{Y_1}(y)\geq\tau\} \nonumber \\
Q^c_{Y_0}(\tau)&=&\inf\{y:F^c_{Y_0}(y)\geq\tau\} \nonumber
\end{eqnarray}
where all notation follows that of section \ref{sscn:QTE}.}

It is worth noting that these methods, while more demanding than standard QTE models given the required conditions of instrumental validity, are potentially well suited to applications in economic history where concerns exist relating to endogeneity.  For example, \citet{Aaronsonetal2020} estimate the impacts of the number of children born to women -- a clearly endogenous variable -- on maternal labour supply, using data over more than two centuries, by leveraging the birth of twins as an IV.  In this case monotonicity assumptions are very likely met given that twin births should have weakly positive impacts on completed fertility.\footnote{Note however that in general, the assumptions necessary for identification with instruments are not trivial.  \citet{BhalotraClarke2019} discuss a number of considerations related to these instruments in a standard linear model.  An entirely different take on IV style models which precluded the LQTE approach described here and allows IVs to generate variation locally in endogenous variables is the work of \citet{Chesher2005,MaKoenker2006}.  This suggests productive ways forward in a quantile framework even if instruments generate shifts at only specific points of endogenous variables of interest.}  Very recent work by \citet{ValenciaCaicedo2021} provides considerable other examples of the use of IV in economic history (though not discussing quantile methods), describing among other examples IVs based on map borders or other geographical features such as rivers, or slopes of terrain, geographic suitability indexes, or Bartik-type instruments.  All such instruments in economic history could be productively introduced into a quantile framework if distributional outcomes are considered.

Finally, prior to turning to applied examples of the use of quantile regression, we note that this particular implementation of IV and endogeneity corrections via LQTE methods is not the only way forward.  Regression discontinuity designs can be similarly cast in this ``Local'' framework, while the instrumental variable quantile regression models (IVQR) of \citet{ChernozhukovHansen2005} provides estimation methods which return average Quantile Treatment Effects (rather than LQTEs), but requiring alternative assumptions.\footnote{In particular, these methods require a rank preservation assumption, restricting the ranks which individuals can take in terms of the ordering of the outcome variable to be the same across differing potential IV assignments.}  These methods are further discussed in \citet{MellyWuthrich2018} and \citet{Chernozhukovetal2018} respectively, and are likely to more productive ways forward when non-binary endogenous variables are considered--in which case LQTE models are less appropriate--or in cases when broader \emph{non}-local QTEs are desired, at the cost of alternative assumptions.  A valuable review of these methods is provided in \citet{ChernozhukovHansen2013}.

\paragraph{Other Extensions and Methods} A range of other contexts which are potentially of use in quantitative studies of economic history can be productively studied in quantile settings.  This includes settings such as longitudinal or panel data and difference-in-difference models, regression discontinuity designs, and non-parametric analyses.  We briefly discuss these settings in turn below, pointing interested readers to relevant references.

Extensions of quantile regressions to panel data following individuals over time have been proposed in \citet{KOENKER2004} where challenges arise given desires to estimate movements across distributions \emph{within} individuals (or panel units) with potentially few repeated individual level data-points.  \citet{KOENKER2004} proposes using a penalized estimator to control for relevant individual-level effects, and much additional work has been conducted to take forward these techniques (a complete discussion is provided in \citet[pp. 11-14]{Lamarche2019}).  Recent work of \citet{GU201968} has suggested using a grouped fixed effect approach, also based on penalized estimators to group similar individuals in a panel setting.  Alternative lines of work, such as \citet{ArellanoBonhomme2016}, suggest viewing (potential individual-level) variation as a problem of unobserved heterogeneity and estimating using an iterative process, super-imposing simulation based estimation procedures on top of standard quantile regression procedures.  A particular setting of interest in cases of longitudinal data consists of the estimation of difference-in-differences models where exposure to some treatment varies within a panel over time such that baseline differences between exposed and unexposed individuals can be captured using pre-treatment periods.  These models have been extended to a quantile setting, see for example \citet{CALLAWAY2018395} for a setting with two time periods and \citet{CallawayTong2019} for a broader panel setting. Computational software is also available to implement these methods in \citet{Callaway2019}

The regression discontinuity design provides credible identification in cases where some dependent variable of interest is moved discontinuously by some arbitrary cut-off.  The use of these methods in economic history has been surveyed by \citet{ValenciaCaicedo2021}, where examples are often based on distance to geographical features or map boundaries.  The regression discontinuity design has been extended to a quantile framework in a very flexible way by \citet{FRANDSEN2012382}, which allows for the estimation of quantile treatment effects `local' to a particular discontinuity, and accompanying computational routines are available to implement this method.

Standard non-parametric regression implementations can capture fully flexible relationships between the mean of some variable $y$ and some independent variable $x$.  Rather than parametric assumptions and linear functional forms (as imposed in equation \ref{eqn:linMod}, the relationship is allowed to vary freely along the support of $x$, allowing for $x$ to have a non-linear impact on the mean of $y$.  This logic can be extended to a quantile regression, if rather than considering a non-parametric relationship between the mean of $y$ and $x$, a non-parametric relationship between specific \emph{quantiles} $\tau$ of $y$ and $x$ are estimated.  This is a particularly flexible way to model heterogeneity which may be well-suited to historical outcomes over which there are few prior assumptions related to the nature of the relationship under study.  A review of the methods is available in \citet[section 3]{Koenker2017}, while computational resources for the implementation of such routines are available in, among others, \citet{Koenker2021,Lipsitzetal2017}.   

\section{Applications in Economic History}
\label{scn:hist}
\subsection{Quantile Regression in Economic History Research}
To have some idea about the extensiveness of the use of quantile regression in economic history, we begin by running a search within the main economic history journals, covering the period from 2000 until the present.\footnote{As we see in the following sub-section, there is even less use of quantile regression prior to 2000.  Two notable exceptions are the studies of \citet{ConleyGalenson1998,ConleyGalenson1994} which provided early illustrations of the power of quantile regression in historical analyses.} Namely, we searched within the Journal of Economic History, Explorations in Economic History, Economic History Review, Cliometrica, and the European Review of Economic History. But we also complemented these searches with searches of other economic history journals as discussed below, estimating that around 50-55 journal articles have made use of quantile regression as an analytical tool during the last two decades, although only a handful of these are dated pre-2005.

The journal which has most frequently published articles employing this technique is the Journal of Economic History, which published 19 articles using quantile regressions for a wide range of topics from 2000, including: Max Weber's hypothesis on the role of Protestantism for economic development \citep{Kersting2020}; the market for paintings in Florence and Italy between 1285 and 1550 \citep{Etro2018}; the interaction between inequality and financial development in the US during the late nineteenth century \citep{Jaremski2018}; fluctuations in technology during the Great Depression in the US \citep{Watanabe2016}; the credibility of fixed exchange rates during the classical gold standard era \citep{Mitchener2015}; and inequality of wealth in the Ottoman Empire \citep{Cosgel2012}; just to mention those published after 2010.  These few examples make clear the broad applicability of quantile methods to questions of interest in economic history, covering issues in micro, macro and financial economics.

Explorations in Economic History has also published articles using quantile regression based methods with some frequency.  A scoping review identified six such article (see also the following sub-section). Namely, \citet{Walker2000}, which explores the degree of economic opportunities in San Francisco compared to other regions around mid-nineteenth century; \citet{Dupont2007}, which tests for contagion in bank runs in Kansas during the panic of 1893; \citet{Canaday2008}, which deals with the relationship between wealth and wealth accumulation by both blacks and whites in South Carolina between 1910 and 1919, and its determinants; \citet{Drelichman2014}, which reconstructs housing costs for various social groups and traces the effect of exogenous shocks on the rental market for Toledo, Spain, between 1489 and 1600; \citet{Alvarez2018}, which deals with the relationship between human capital and male labour earnings in eighteenth-century Spain; and \citet{Callaway2018}, which measures the union wage premium for several US-cities circa 1950, using unconditional quantile methods.

The Economic History Review, in turn, has published five articles where quantile regressions were used: \citet{Temin2008} analyses the cost and availability of private bank credit between 1702 and 1724; \citet{Gazeley2011}, in turn, estimates urban poverty among working families in the British Isles circa 1904; \citet{Brown2018} deals with the causes of fluctuations in infant mortality rates in Bavaria during the 1820s-1910s; \citet{Artunc2019} examines the composition of firm ownership and entrepreneurship in Egypt between 1910 and 1949; while \citet{Karagedikli2021} estimates real hedonic house prices and urban wealth inequality for the housing market between 1720 and 1814 in the Ottoman Empire.  Note that as above, these articles show both a broad scope of themes, as well as a broad scope of geographic and temporal settings which have been productively analysed with these models.
	
The European Review of Economic History has also published five articles making use of quantile regression: \citet{Koepke2005}, which provides the first anthropometric estimates of the biological standard of living in Europe during the first millennium AD; \citet{Dincecco2009}, which performs a statistical analysis of political regimes and sovereign credit risk in Europe from 1750 to 1913; \citet{Dribe2009}, who analyses the importance of demand and supply factors in the Swedish fertility transition between 1880 and 1930; \citet{Kholodilin2016}, that analyses the housing rental dynamics of Berlin during World War I; and more recently, \citet{Jorge-Sotelo2020} focused on the impact of currency depreciation on international capital flows in Spain between 1928 and 1931 crisis.

Cliometrica has published six articles using quantile regressions, the first of these less than a decade ago: \citet{Carson2012}, compared body mass index values of late 19th- and early 20th-century amongst African-Americans groups (i.e. blacks versus mulatto); \citet{Ogasawara2015}, which deals with the impact of social workers on reducing infant mortality rates in inter-war Tokyo; \citet{DuPlessis2015}, who for the period 1700-1725, estimated hedonic slave price indices and the value of their marginal productivity; \citet{GonzalezVal2017}, who analysed the impact of market potential on the structure and growth of some Spanish cities during 1860–1960; \citet{Ogasawara2019}, which deals with the treatment effects of piped on diseases in industrializing Japan (1920s-1930s); and finally, \citet{Keywood2020}, who tests the relationship between elite numeracy and elite violence in Europe from 500 to 1900.
	
We also run searches in other economic history journals, but where quantile regression was observed to be very infrequently used. For example, in the main Spanish economic history journal, Revista de Historia Economica -- Journal of Iberian and Latin American Economic History, quantile regressions were used in only four papers; in the Australian Economic History Review in only one article; in the Economic History of Developing Regions, similarly in only one article; while in the Scandinavian Economic History Review it has appeared twice. Finally, in the three main business history journals (Business History, Business History Review and Enterprise \& Society), it was used in three articles, twice at Business History and once at Enterprise \& Society, although this is rather unsurprising since this sub-discipline cultivates a less quantitative approach to history.

\subsection{The Potential for Quantile Regression  Methods in Economic History}
Reviewing cases where quantile regression \emph{is} used suggests the broad applicability of these methods across themes and settings in economic history, but does not indicate the additional scope for use in papers which focus principally on mean effects.  To see the \emph{potential} for Quantile Regression, we carried out a review based on all papers published over the last 30 years in a specific economic history journal (Explorations in Economic History, hereafter EEH).  This allows us to gain a more complete picture of trends in published research in a specific highly-cited economic history journal, consider the methods used, and the potential for distributional analyses in economic history.  In this review, we read each paper published over the last 3 decades (the first volume of 1990 to the last volume of 2020), and classified each paper according to whether it contained empirical methods, or was theoretical.  In the case of empirical papers, we then classified these by method (OLS, Logit, Quantile Regression, and so forth) and finally whether principal dependent and independent variables are continuous, discrete, ordinal, etc. 

\begin{figure}[htpb!]
\caption{Trends in the dependent variable in papers published in \emph{Explorations in Economic History}}
\label{fig:area}
\subfloat[Large groups\label{fig:aggregated}]{%
\includegraphics[scale=0.57]{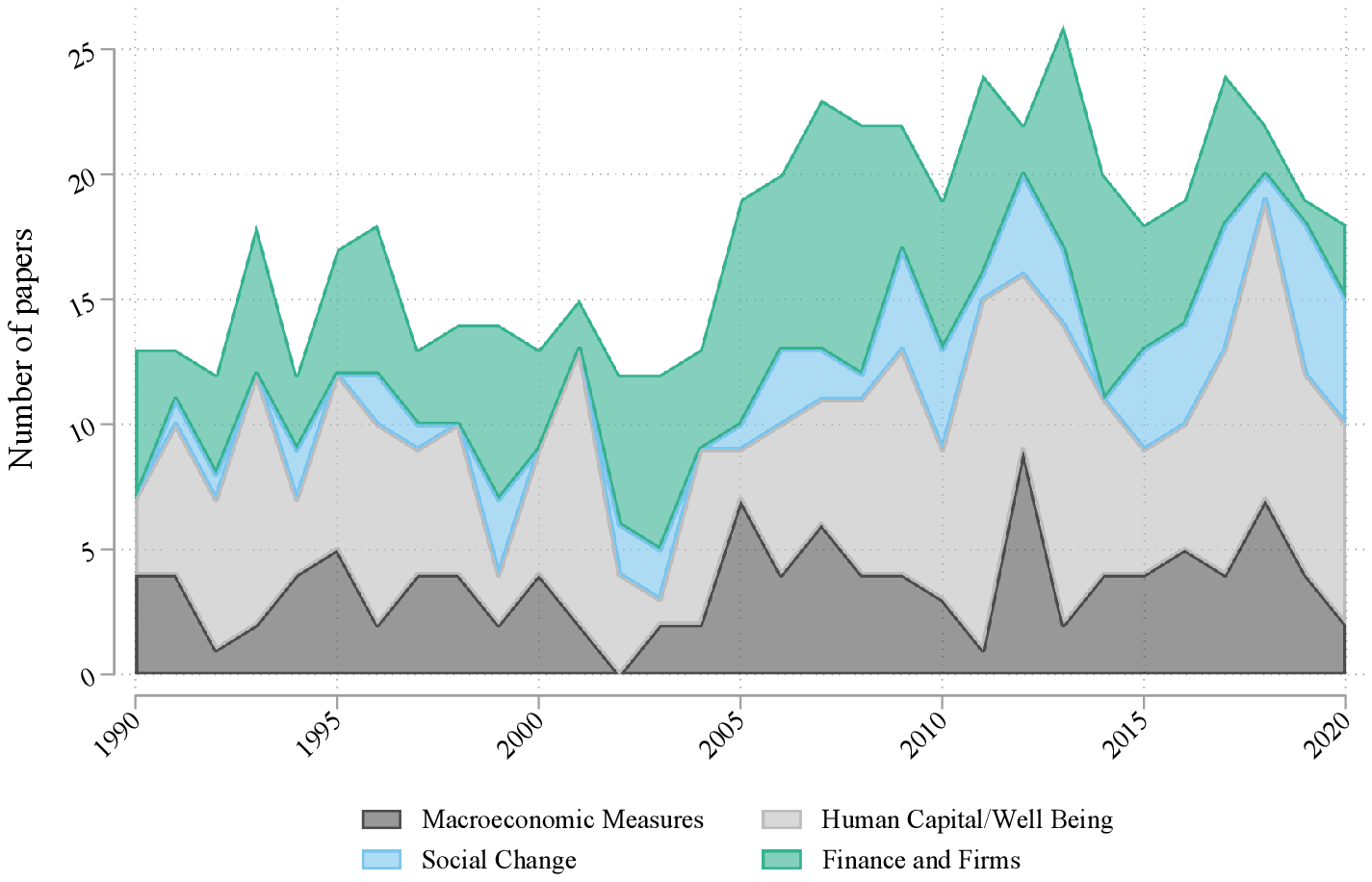}%
}
\subfloat[Small groups\label{fig:disaggregated}]{%
 \includegraphics[scale=0.57]{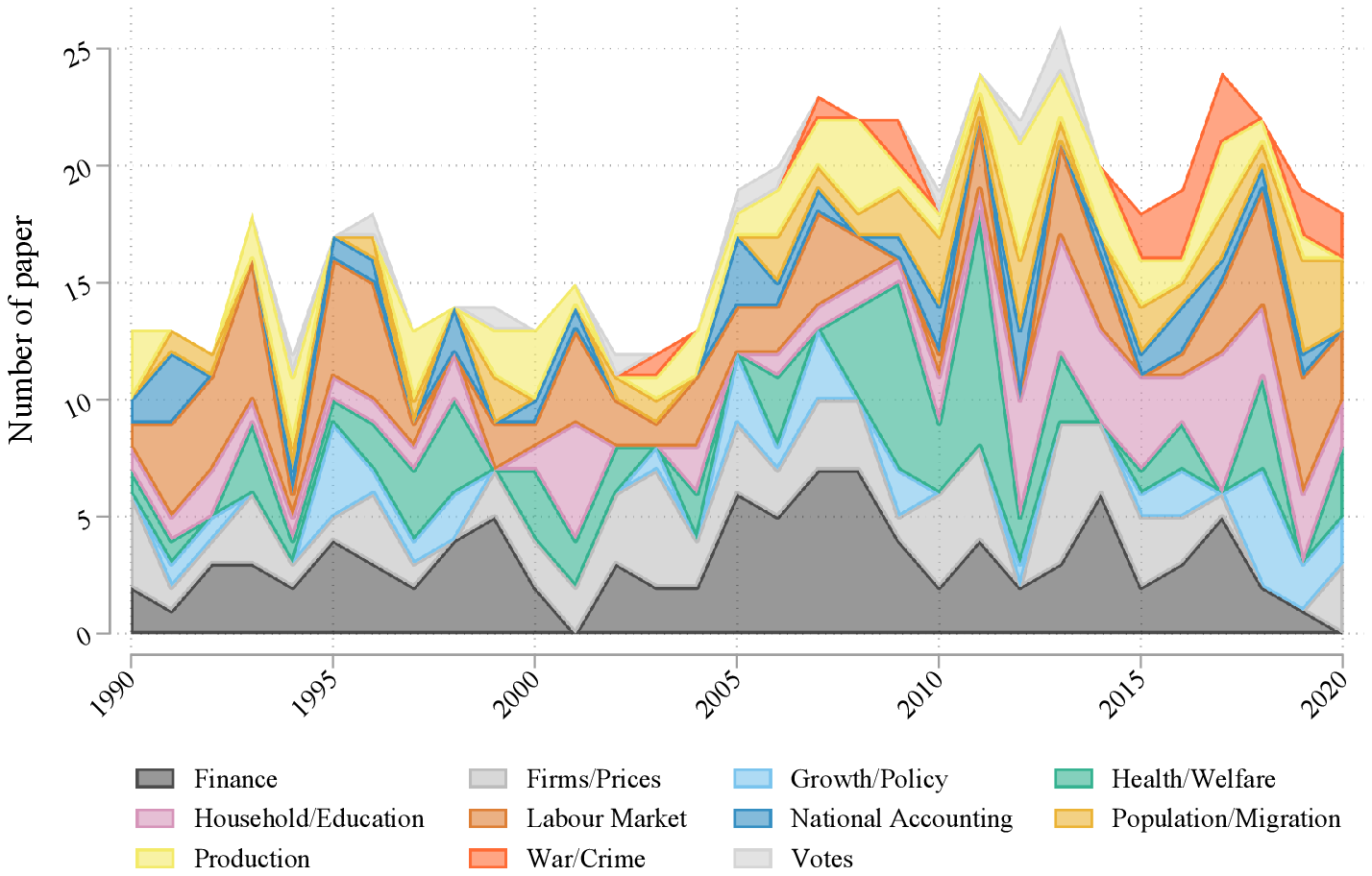}%
}
\floatfoot{\textsc{Notes:} Based on our reading of each paper published in EEH we classify each paper as belonging to a particular (single) area, as described in the legend in panel (b).  These classes are further aggregated in panel (a), where `Macroeconomic Measures' refers to Growth/Policy, National Accounting and Production; `Human Capital/Well Being' refers to Health/Welfare, Household/Education and Labour Market; `Social Change' refers to Population/Migration, War/Crime and Votes; and Finance and Firms refers to Finance and Firms/Prices.  A file summarising all papers in terms of a number of characteristics discussed in this review is provided as an online appenix to this paper.} 
\end{figure}

Prior to considering trends in measures and methods, it is useful to provide descriptive evidence of the key themes studied over the past 30 years in this particular journal.  In Figure \ref{fig:area}, we plot the broad areas which papers study, as classified based on our reading of the literature. These classifications refer to the dependent variable under study in all empirical papers published in EEH. The right-hand plot classifies papers into one of four aggregate areas: macroeconomic measures, human capital and well-being, social change (which includes papers studying topics such as conflict, migration, and changing political landscapes), and finance and firms.  In general we note the largest of these groups is that related to micro-economic areas such as education, health and labour markets (the shaded light gray area), followed by themes related to finance and firms, with relatively less work focused principally on macroeconomic indicators.  In the left-hand panel we provide more disaggregated classes, seeing more noisy patterns, and evidence of broad coverage of topics within economic history.

While it is illustrative to observe trends in the published record in economic history, here we are most interested in considering how the types of questions and measures used in these papers relate to their potential for using quantile methods.  At minimum, this implies that dependent variables need to be continuous, given that quantile methods examine impacts of some independent variable across the distribution of the dependent variable.  

\begin{figure}[htpb!]
\caption{Trends in Publications in an Economic History Journal}
\label{fig:trends}
\subfloat[All Papers\label{fig:allPapers}]{%
\includegraphics[scale=0.57]{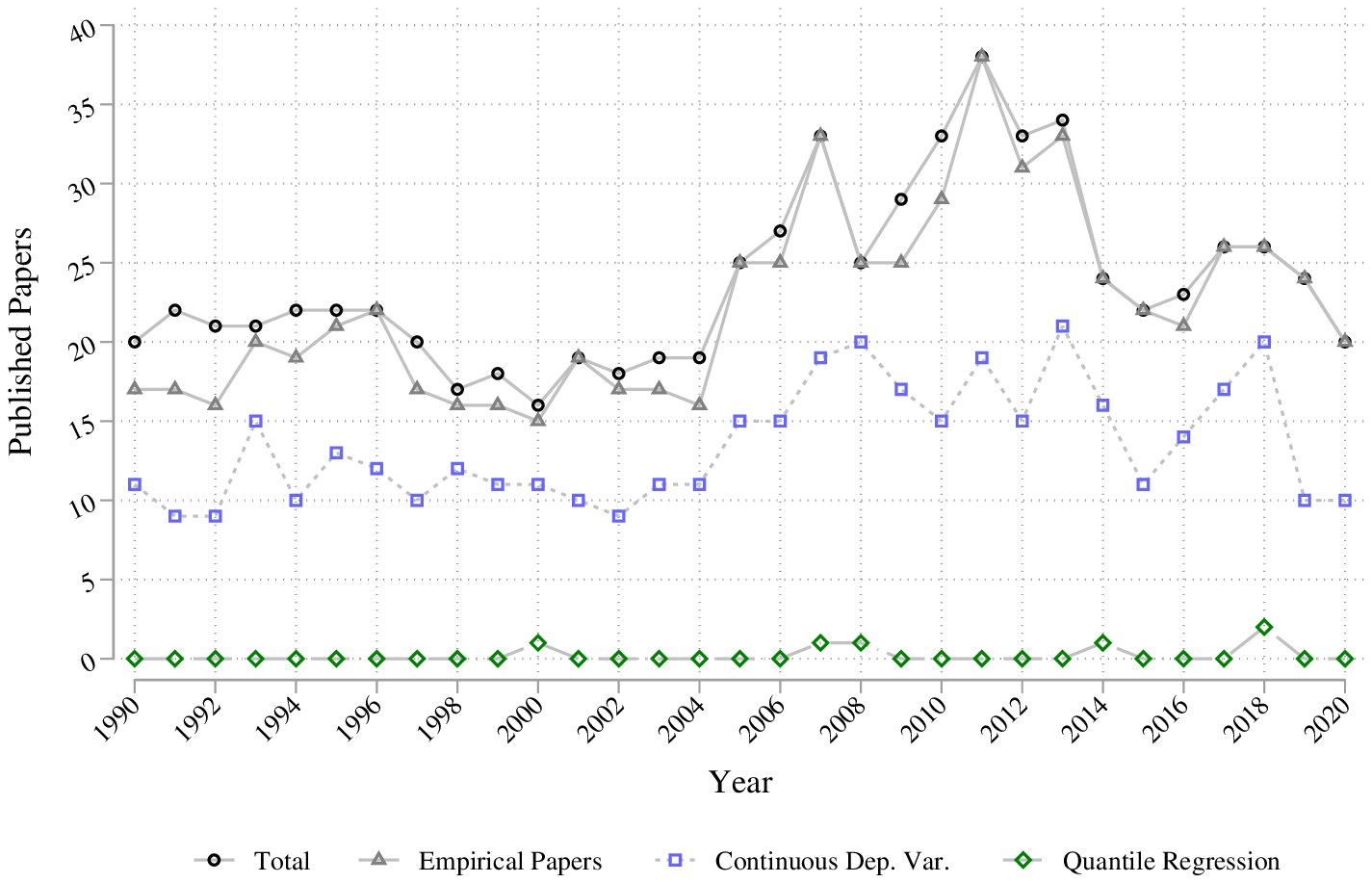}%
}
\subfloat[Proportion of Papers\label{fig:propPapers}]{%
 \includegraphics[scale=0.57]{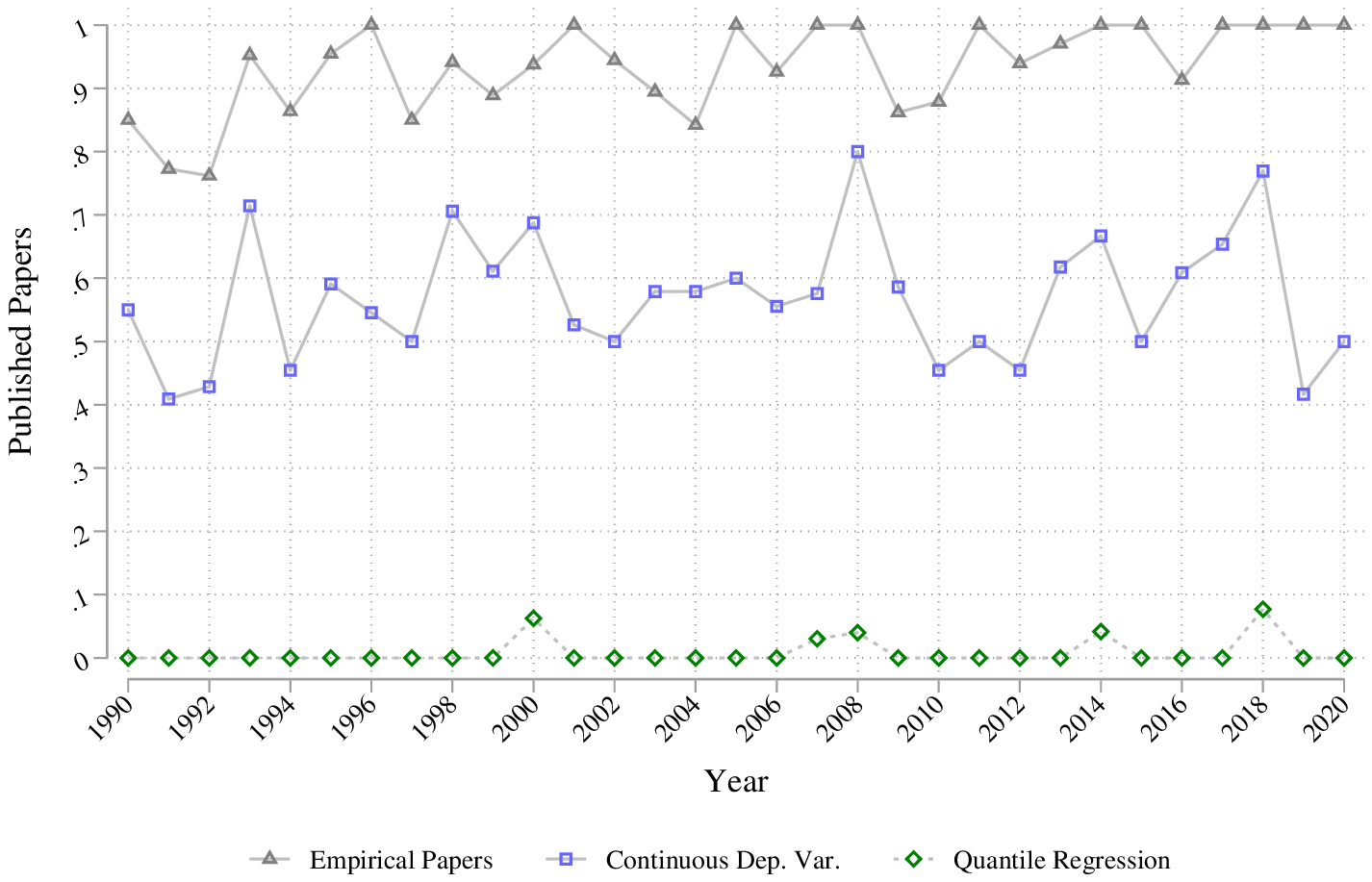}%
}
\floatfoot{\textsc{Notes:} The total number of papers (panel A) and the proportion of all papers (panel B) refer to all papers in the journal EEH, based on the isssue in which they are published.  This data was collected based on a reading of each paper and classification of the methods and `principal' dependent variables.} 
\end{figure}

Figure \ref{fig:trends} plots the number and share of papers published in \emph{EEH}, classifying them by whether they are empirical or not, and in the case of empirical papers, whether the dependent variable is continuous, and hence suitable for quantile analyses.  Finally, we plot the number and share of papers that effectively do undertake quantile analyses.  In Figure \ref{fig:trends} panel (a) we can see how the amount of total articles published in \emph{EEH} was around 20 annual publications, with an increase to around 30 from 2005 onwards. We can see how the number of empirical papers follows a similar trend to that of total articles, and corresponds to 93.1\% on average of the number of total articles. In recent years, this value is even higher, with 100\%, or close to 100\% of articles containing at least some considerable empirical portion, pointing to the growing frequency of empirical evidence in studies in economic history.  

While this is illustrative of the importance of quantitative analyses in understanding historical processes in economics, it does not necessarily imply a suitability for analysis with quantile methods.  However, if we observe the articles that are based on continuous dependent variables and hence amenable to quantile-based methods, we observe that they correspond to about 56.5\% of the quantity of total articles (over the past 30 years).  Despite these two stylised facts: (i) that there is considerable quantitative evidence brought to bear in economic history papers, and (ii) that many of these analyses are based on variables which have continuous distributions and hence potentially interesting distributional features, we observe little use of quantile methods.  Among the 738 papers examined, 692 of which were empirical, only 6 studies used quantile regression (as described in the previous sub-section).  Unless there is a particular interest in the mean, these facts taken together point to a considerable margin for the adoption of other estimation methods such as quantile regression or QTE methods within economic history.

\begin{table}[htpb!]
  \centering
  \caption{Classification of Dependent and Independent variables}
  \label{tab:codvar}
    \scalebox{0.84}{
    \begin{tabular}{llcccc}\toprule
     &&\multicolumn{2}{c}{Dependent}&\multicolumn{2}{c}{Independent} \\
      \cmidrule(r){3-4}\cmidrule(r){5-6}
      & Type            &Number&Proportion&Number&Proportion\\ \midrule
      & Continuous      & 418  & 76.70 & 207 & 37.91\\
      & Binary          & 58   & 10.64 & 47  & 8.61\\
      & Discrete        & 23   & 4.22  & 12  & 2.20\\
      & Ordinal         & 3    & 0.55  & 0   & 0\\
      & Multiple        & 10   & 1.83  & 115 & 21.06\\
      & Continuous-Binary   & 19 & 3.49& 77  & 14.10\\
      & Continuous-Discrete & 9  & 1.65& 21  & 3.85 \\
      & Continuous-Dummy    & 0  & 0   & 25  & 4.58 \\
      & Continuous-Ordinal  & 1  & 0.18& 0   & 0  \\
      & Dummy-Dummies       & 0  & 0   & 30  & 5.49 \\
      & Discrete-Binary     & 4  & 0.73& 12  & 2.20 \\
      \bottomrule
      \multicolumn{6}{p{11.8cm}}{{\footnotesize Notes: Multiple classification is article with more than two types of variables. The classification that presents 2 types of variables, correspond to papers that present more than one regression in which its dependent variables are continuous plus one or more variables of another type, and both regressions are relevant in the research.}}
  \end{tabular}}
\end{table}

Finally, before providing a simple illustrative example of patterns observable in quantile regressions in a particular study of historical economic processes, we note how these continuous dependent variables are situated within analyses.  In Table \ref{tab:codvar} we provide a full break down of all the classes of dependent variables observed across all studies from EEH.  418 papers had a single continuous outcome variable of interest, while another 29 had a continuous outcome variable as well as other non-continuous measures.  Of these 447 studies, 201 had independent variables which were also continuous, and so more suited to quantile regression methods, while 33 had a binary independent variable of interest, and so are potentially well suited to QTE methods.  Many other studies have various types of independent variables, and as such either QTE or quantile regression methods may be appropriate.


\section{An Illustrative Example Based on 19\textsuperscript{th} Century Demographics and Economic Growth}
\label{scn:example}

\begin{figure}[htpb!]
\caption{Distributions of individual height and sub-national rates of historical growth}
\label{fig:hist}
\subfloat[Adult height in cm\label{fig:height}]{%
\includegraphics[scale=0.55]{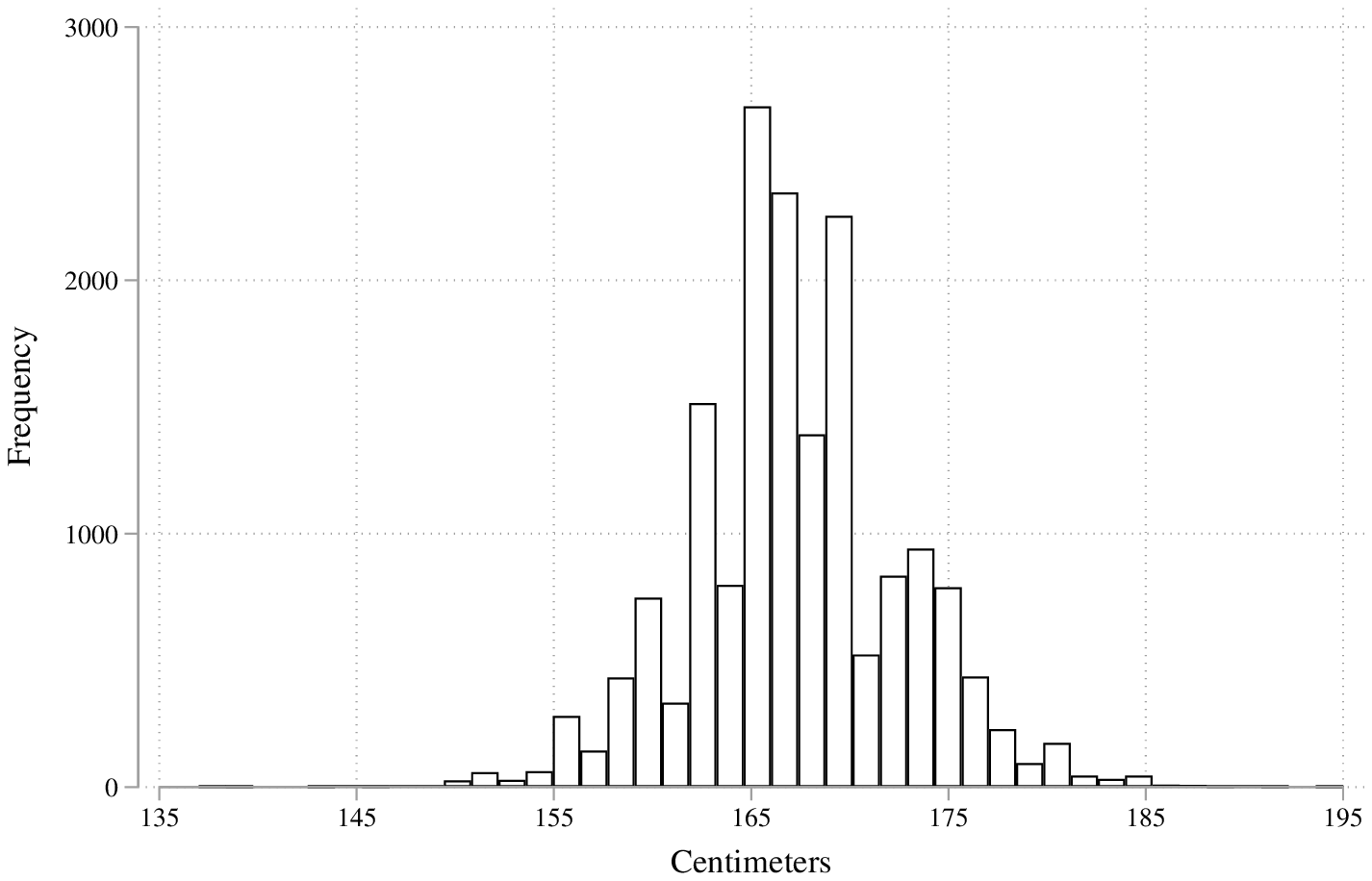}%
}
\subfloat[Growth rate over all decades\label{fig:growth}]{%
\includegraphics[scale=0.55]{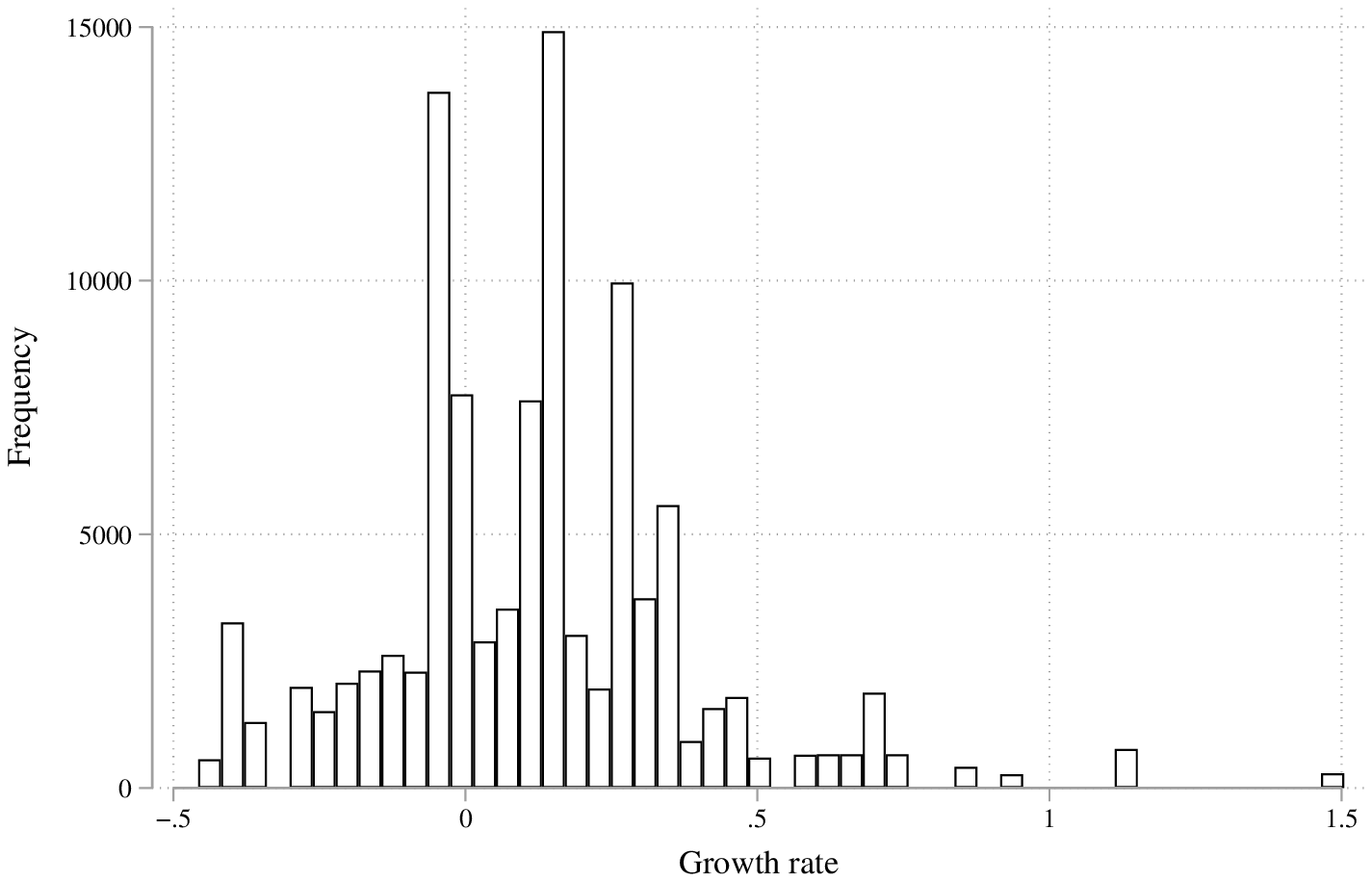}%
}
\floatfoot{\textsc{Notes:} Histogram: Panel (a) distribution of adult male height in Chile, 1890s–1940s (in centimetres, 17,293 observations); Panel (b) distribution of growth rate in Chile, 1890s–1940s.\\ Source of data: Panel (a) \citet{llorcajanaetal2019,llorcajanaetal2021}, Panel (b) \citet{badiamiro2008}.}
\end{figure}

As a brief illustration of the use of quantile regression, and how it can shed light on patterns across the distribution of outcomes which are hidden by standard analyses, we consider the empirical setting described in \citet{llorcajanaetal2019,llorcajanaetal2021}. As laid out there, we gathered information on all the 36,371 records of Military Personnel born in the 20\textsuperscript{th} century and 3,283 record of Military Personnel born in the 19\textsuperscript{th} century in Chile.\footnote{There are many historical precedents to the study of height in economic history, although less work extending to quantile analyses. Among many other references \citet{WachterTrussell1982} discuss historical measurement of height, though work goes back much further, for example Quetelet's discussion of classifying populations.} These records relate mainly to soldiers or low-ranking officers (the Army's Historical Archive). Full information on this process and these data are available in \citet{llorcajanaetal2019,llorcajanaetal2021}.  Of those individuals, we generate a final database of the Chilean-born individuals aged between 17-55, which here we cross with rates of economic growth in the province in which the individual was born. These rates of economic growth are calculated from historical evidence collected by \citet{badiamiro2008}, which is the best sub-national evidence of economic conditions in Chile, available covering the periods of 1890-1950. From these data sources, we are able to combine a large micro-level sample of human height as well as measures of economic growth by decade and province of birth. Finally, we have a database of 17.293 individuals. The reduction of the database is explained given the data availability of growth rates (1890-1940).\footnote{We note that the sample reduction from the initial 36,371 digitized records owes to the availability of historical measures of growth.  We calculate growth records in each decade as: \[\frac{GDPpc_{p,t+1}-GDPpc_{p,t}}{GDPpc_{p,t}},\] where $p$ indexes provinces and $t$ indicates decades, and as such this will not be defined in the decade of 1950, given that comparable records are not available for 1960.  We thus limit the final estimation sample to all individuals born in the 60 years between the 1890s and 1940s.}  
Descriptive plots of these data on height and exposure to economic growth are provided in Figure \ref{fig:hist}, suggesting considerable variation in observed heights, and also in changes in economic conditions within provinces over time.

To document these methods, we consider a particular model seeking to determine the effect of economic growth during an individual's formative years on their adult height.  Relationships between height and economic development have been discussed in the past, including over long periods, such as \citet{Peracchi2008}'s study of Italian height from the 1730s-1980s.  Here, we are interested in a model of the following type:
\begin{equation}
\label{eqn:heightReg}
\text{Height}_{ipt} = \beta_0 + \beta_1 \text{Growth}^0_{pt}  + \beta_2 \text{Growth}^6_{pt} + \beta_3 \text{Growth}^{12}_{pt}  + \beta_4 \text{Growth}^{18}_{pt} + \mu_p + \lambda_t + \varepsilon_{pt},
\end{equation}
where the height of an individual $i$ born in province $p$ and year $t$ is regressed on the growth rate in that province when the individual is born (Growth$^0$), when they are aged 6 years (Growth$^6$), when they are aged 12 years (Growth$^{12}$) and when they are aged 18 years (Growth$^{18}$).  It is important to note that given challenges in collecting data on economic patterns around 200 years in the past, these measures of growth are the best available estimates, but should be considered as noisy measure of the economic conditions during an individual's growing years.\footnote{We also note that we consistently use measures related to an individual's province of birth, given that from Military data we know where they are born.  However, in the case of individuals moving between provinces in the country, these measures are noisy proxies of exposure to local economic conditions.}  This model includes province and decade fixed effects ($\mu_p$ and $\lambda_t$ respectively), capturing idiosyncratic regional or temporal factors, such that measures of growth are not simply proxying regional or time-specific factors that correlated with height.

\begin{figure}[htpb!]
\caption{Quantile Regression}
\label{fig:qreg_hat}
\includegraphics[scale=1.1]{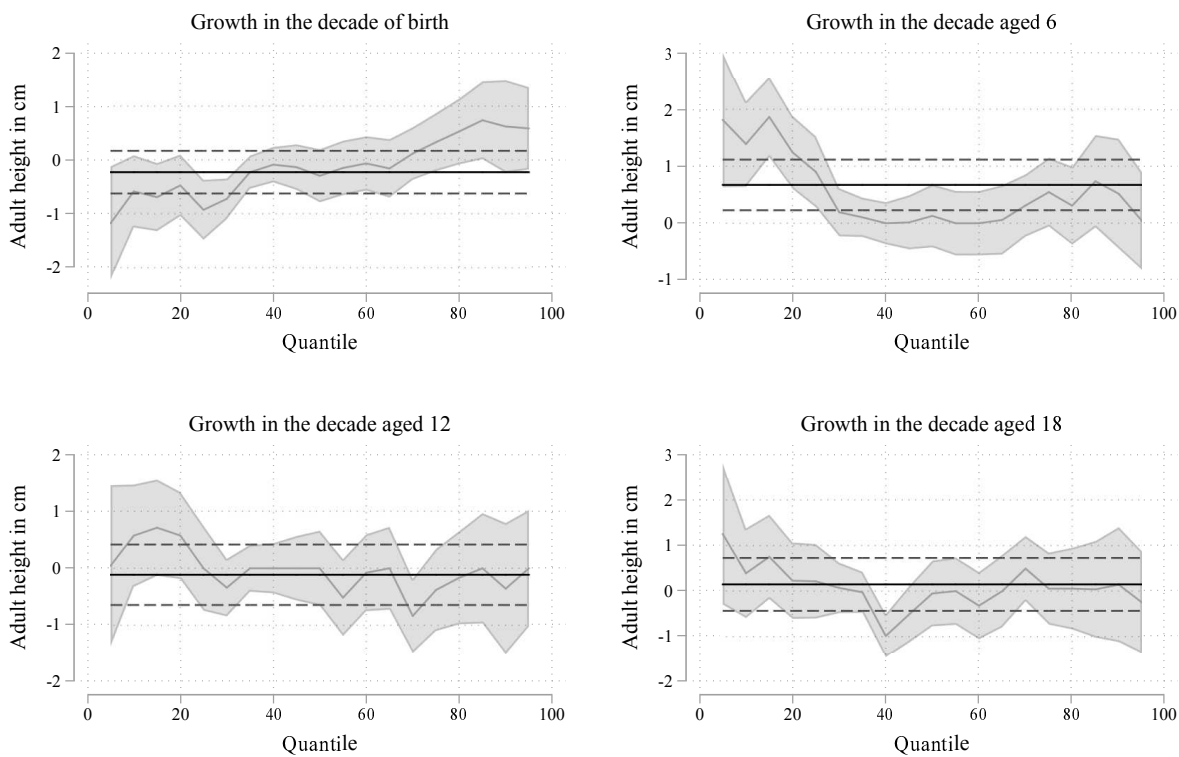}%
\floatfoot{\textsc{Notes:} Each sub-plot presents the OLS and quantile regression estimates of the effect of local rates of economic growth at particular points of an individual's life on their adult height. Solid black lines and dashed lines refer to OLS point estimates and 95\% CIs, while gray solid lines and shaded areas refer quantile regression point estimates and 95\% CIs.} 
\end{figure}

To consider what can be gained from estimating quantile regression in this particular setting of interest in economic history, we present estimates from equation \ref{eqn:heightReg} by both standard OLS, and using quantile regression following equation \ref{eqn:qreg} at quantiles $5, 10, \ldots, 95$.  These results are displayed in Figure \ref{fig:qreg_hat}, where we plot estimates on each of the parameters $\beta_1, \beta_2, \beta_3$ and $\beta_4$ on average (OLS, indicated by the solid horizontal black line), and across the distribution (quantile regression, indicated by the solid gray line).  95\% confidence intervals are indicated by dashed lines (OLS) or shaded intervals (quantile regression).  

The top-left panel documents a null effect of exposure to growth in the decade of birth on adult height when considering average effects, although marginally negative effects lower in the distribution of adult height, and marginally positive effects higher in the distribution.  Where results are most striking is in the top right-hand panel.  While exposure to growth in the decade in which a child is aged 6 years is observed to  significantly increase adult height (OLS estimates suggest a 1\% higher rate of growth is associated with nearly 1cm in additional height), this effect is very heterogeneous, with large impacts at the lower end of the distribution of height, and null impacts among taller individuals.  Here the value of considering quantile regression is clear.  These results point to a particular value of economic growth in human condition, which -- at least in this early life period where children are growing significantly -- is most relevant among those who are shorter.  This is significant, as adult height is a well known marker of health, suggesting that economic growth during this period of childhood in this context has done the most to pick up individuals who have the worst health stocks.  While some similar patterns may even persist when children are older (bottom panels), these effects are observed to be most notable in the early years of life when children are particularly sensitive to their conditions.  While these results are descriptive, and presented in part to illustrate the utility of quantile regression in research in economic history, they are nonetheless able to point to key distributional factors which are of relevance in understanding human demographics and sensitive periods of human capital accumulation, as well as the value of historical periods of growth -- beyond just population averages.



\section{Conclusions and Ways Ahead}
This paper seeks to provide a preview of a range of empirical methods which are relevant to consider impacts of some independent variable(s) of interest, across the entire distribution of a continuous dependent variable of interest.  We specifically seek to describe these methods, and motivate their adoption more widely in literature in economic history where considerable work is often spent to collect rich (continuous) outcome measures of interest.  These methods can thus contribute to fully taking advantage of such data collection processes or existing data respositories which have been collected based on considerable efforts in collating, systematising, or digitising historical records.

We discuss both standard quantile regression methods originally laid out by \citet{KoenkerBassett1978} and also document how these have been fruitfully applied in a more recent ``treatment effects'' literature, with the application of Quantile Treatment Effects.  We discuss a number of other extensions which may be of particular interest for researchers in economic history such as quantile regression with measurement error, additional ways to loosen parametric assumptions, and potential solutions to endogeneity in these models.  This is a very large and ever-growing literature in econometrics, and so here, while aiming to provide a broad overview of the field, we do not claim to comprehensively survey the entire field nor the full depth of all models.  Fortunately, there are a number of full textbook or handbook references such as work of \citet{KOENKER2004,Koenkeretal2017}, to which we point interested readers in cases where a more comprehensive econometric base of these models is desired.

While we argue that these models are well suited to research in economic history, we suggest that there is scope for considerably more work in this line.  Fortunately, these methods are accompanied by a range of computational tools which mean that their application can be viewed as part of a quite standard toolbox for interested practitioners.  In closing, it is worth pointing to the functionality of these packages, which are significant and generally open source contributions of the methods discussed in this paper, allowing for these methods to be adopted at relatively low cost.

Computational languages widely used in economics and economic history such as R, Stata, Julia, Python, MATLAB and so forth generally all have a standard implementation of quantile regression allowing for simple implementation of these models.  For example, \citep{Koenker2021} provides the R \texttt{quantreg} package which contains (among \emph{many} other things) a standard quantile regression interpretation as \texttt{rq}, while Stata's \texttt{qreg} provides a stable option for both estimation and various inference procedures.  However, many extensions to these commands' `standard' procedures are available, including packages to extend analyses to quantile treatment effects such as \texttt{ivqte} in Stata \citep{FrolichMelly2010} or the \texttt{QTE} package in R \citep{Callaway2019}.  Each of these QTE libraries extends in numerous ways to cases where treatment assignment is endogenous, and, in the case of \texttt{QTE} to difference-in-differences models \citep{Callaway2019}.  Within the `universe of \texttt{quantreg} in R there are many other extensions, including a wide range of inference procedures, non-linear models, LASSO models, and plotting functions.  Given the highly applied nature of these methods, and the considerable recent extensions in the field, many new papers are also accompanied by computational code -- most frequently in R or Stata -- including the examples discussed in the paper such as recent advances in selection models \citep{Siravegna2020,BE2020}, and quantile analyses in regression discontinuity designs \citep{FRANDSEN2012382}.  All in all, these models present an extremely flexible, accessible and extendable series of analytical routes to researchers in economic history, and should be viewed as a key component of analyses, allowing for a much richer consideration of the distributional effects of historical phenomena in economic processes.



\clearpage
\end{spacing}
\bibliography{quantBib}



\end{document}